# Finite element modelling of immiscible two-phase flow in oil reservoirs

Taofik H. Nassan[a]* and Mohd Amro[a]

[a]*Institute of Drilling Engineering and Fluid Mining, Freiberg University of Technology, Freiberg, Germany*

*Corresponding author

ORCiD:

Taofik H. Nassan

https://orcid.org/0000-0002-3680-4268

# Finite element modelling of immiscible two-phase flow in oil reservoirs

Abstract

Reservoir simulators utilize numerical techniques to solve the governing equations of fluid flow in porous media and they are essential tool for oil and gas fields' development. In practical reservoir simulation, the finite difference method (FDM) is the common numerical technique owing to its simplicity in application. In this paper, we introduce the finite element method (FEM) as a numerical tool to simulate an example in 3D and prior to that, Buckley-Leverett problem is used to validate using a commercial FEM-based software package that is applied through this paper. To achieve the proposed target, the mathematical model of the immiscible-two phase problem is reviewed and formulated to be applied easily in the proposed software. Immiscible two-phase flow in 3D is simulated on the 1/4$^{th}$ inverted five-spot benchmark and the results are shown and discussed. The results show that the FEM can be used efficiently to simulated immiscible two-phase flow in oil reservoirs with an acceptable CPU time and this is due to faster solvers' advances in the last few years.

## 1. Introduction

Reservoir simulation is an inevitable tool for reservoir engineers to study and predict the reaction of reservoir to different enhanced oil recovery schemes, let alone history matching to match measured data to production history. The underlying equations relevant to reservoir simulation are generally typical conservations laws with some extra physics included according to the process that has to be simulated. These laws are manifesting themselves in a PDE form and to solve a set of PDE system for a field-scale reservoir, these equations must be solved numerically. Numerical methods are the only techniques to discretise these equations and convert them into a set of algebraic equations which are amenable to be solved by direct or an iterative solver. The main numerical techniques that are used in numerical computation classified in the order of increasing accuracy and the cost of computation are: finite difference method (FDM), Finite volume method (FVM), and Finite element method (FEM), respectively. Most of commercial and research simulators depend heavily on FDM as a numerical technique to discretize the governing equations of different physics inside the reservoir (Aziz and Settari 1979; Ewing 1983; Ertekin, Abu-Kassem and King 2001; Chen, Huan and Ma 2006; Chen 2007; Abou-Kassem, Islam and Farouq Ali 2020). The first models for two-phase flow that were developed to imitate displacing oil by water in the laboratory experiments used FDM to solve the coupled equations for pressure and saturation (Douglas, Peaceman, and Rachford 1959). Later on, more complex models were introduced to combine as much physics as they can (Gottfried 1965; Coats 1968; Shutler 1969; Coats, George, and Marcum 1974; Allen 1985; Chen, Huan and Ma 2006). The complexity of the models were walking side by side with computer capabilities. Although the FDM method is a quick and moderately robust technique, it has limitations in solving strongly nonlinear equations or domains with complex geometry.

Another numerical method applied in structural analysis, heat transfer and fluid flow, is the finite element method (FEM). The FEM performs better than FDM in handling nonlinear behaviour, modelling irregularly shaped bodies, and in incorporating various boundary conditions. The advantages of FEM has made it much popular choice in different engineering applications. Although researchers have investigated the use of FEM for reservoir simulators since the 1980$^s$ (Chavent **and** Jaffre, 1986), the high computational expense in large scale FEM simulation has constrained the implementation of the method in petroleum reservoir engineering. However, the advancement in high-performance computing (HPC) and the development of more efficient solvers in recent years can help to gain the potential advantages of FEM in reservoir engineering.

In this paper, modelling of immiscible two-phase flow in oil reservoir using FEM is the ultimate target. However, to validate using the proposed software (Comsol Multiphysics®), 1D Buckley-Leverett problem is solved firstly and the results are illustrated. Then the $1/4^{th}$ inverted five-spot benchmark is used to test the applicability of the software to simulate 3D model and check its efficiency.

While usually commercial reservoir simulator is a black box package, Comsol Multiphysics® offers the possibility to use what is called equation based module (Comsol 2019; Nassan 2018; Nassan and Amro 2019; Nassan and Amro 2020; Onaa, Nassan and Amro 2020). In this module one can input the desired equations in different forms and apply the initial and boundary conditions of the problem. This is why this kind of modelling can be coined as "open box simulation package". It needs some experience to formulate the mathematical model of the problem and an experience in applying the initial and boundary conditions and also tuning of the model is very important. Another advantage that can be used in this package also is applying the equation system in weak form which is a very advanced method in modelling and can be discussed in a future paper.

The rest of this article will be organized as follows. Chapter 2 reviews the mathematical modelling of immiscible two-phase flow then validation of the software package using Buckley-Leverett in 1D is introduced in chapter 3. Two-phase flow in 3D will be demonstrated in chapter 4 before the discussion in chapter 5. Finally the conclusions of the paper will be drawn.

## 2. Mathematical modelling of immiscible Two-phase flow

Two-phase flow in any porous medium can be described by continuity and momentum equations for each phase (Chen et al., 2006; Chen, 2007; Blunt 2017).

Mass conservation (continuity equation):

Let the porous medium fills a domain $\Omega \subseteq \mathbb{R}^3$. Conservation equation for each phase $\alpha$ is written as:

$$\frac{\partial(\varphi S_\alpha \rho_\alpha)}{\partial t} + \nabla \cdot (\rho_\alpha u_\alpha) = q_\alpha \; ; \; \alpha = w, o \qquad (1)$$

Darcy's law (momentum equation):

It must be defined for each phase $\alpha$ and reads:

$$u_\alpha = -\frac{\boldsymbol{k} k_\alpha}{\mu_\alpha}(\nabla p_\alpha - p_\alpha g \nabla z) \; ; \; \alpha = w, o \qquad (2)$$

Also the mobility for each phase can be defined as:

$$\lambda_\alpha = k_{r\alpha}/\mu_\alpha$$

Where total mobility is $\lambda = \sum_\alpha \lambda_\alpha$

Fractional flow for each phase is $f_\alpha = \lambda_\alpha/\lambda$ and $\sum_\alpha f_\alpha = 1$

Total velocity in the pore space is the sum of the velocity of both phases, thus it can be written as:

$$u = u_o + u_w \qquad (3)$$

To close the equations' system, two customary equations for saturation and capillary pressure are introduced as follows:

$$S_w + S_o = 1 \qquad (4)$$

$$p_c = p_o - p_w \qquad (5)$$

## *2.1 Fractional flow formulation of immiscible two-phase model*

Continuity equation (1) can be simplified by the following assumptions (Chen, 2007; Abu Kaseem et al. 2020):

- The temperature is constant in the domain;
- There are only two phases: water (w) and oil (o);
- The fluids and the rock (the matrix) are incompressible;
- There are two components: water (w), only in the water phase, and oil (o), only in the oil phase;
- The solid matrix is not poroelastic; meaning that the available pore space (porosity) is constant;

The system of immiscible two-phase flow can be described by two equations and they are referred to as pressure equation and saturation equation and they can be derived from the conservation laws that has been introduced in the previous chapter.

Apply equation (2) into (1) yields:

$$\frac{\partial(\varphi S_\alpha \rho_\alpha)}{\partial t} + \nabla \cdot \left(\rho_\alpha \left(-\frac{kk_\alpha}{\mu_\alpha}(\nabla p_\alpha - \rho_\alpha g \nabla z)\right)\right) = q_\alpha \; ; \; \alpha = w, o \qquad (6)$$

Applying above assumptions and adding oil and water continuity equations yields pressure equation (7):

$$\nabla \cdot u = q_o + q_w \qquad (7)$$

And saturation equation (8):

$$\varphi \frac{\partial(S_\alpha)}{\partial t} + \nabla \cdot u_\alpha = q_\alpha; \; \alpha = w, o \qquad (8)$$

One can obtain the total velocity and phase velocities using equations (2), (3) and (5) in the following form:

$$u = -k[\lambda \nabla p - \lambda_w \nabla p_c - (\lambda_w \rho_w + \lambda_o \rho_o)g\nabla z] \qquad (9)$$

$$u_w = f_w u + k\lambda_o \, f_w \nabla p_c + k\lambda_o f_w \left[(\rho_w - \rho_o)g\nabla z\right] \qquad (10)$$

$$u_o = f_o u - k\lambda_w \, f_o \nabla p_c + k\lambda_w f_o \left[(\rho_o - \rho_w)g\nabla z\right] \qquad (11)$$

In this article capillary forces will be not considered. This leads the capillary pressure to be cancelled from equations (8), (9) and (10):

$$u = -k[\lambda \nabla p - (\lambda_w \rho_w + \lambda_o \rho_o)g\nabla z] \qquad (12)$$

$$u_w = f_w u + k\lambda_o \, f_w [(\rho_w - \rho_o)g\nabla z] \qquad (13)$$

$$u_o = f_o u + k\lambda_w f_o\left[(\rho_o - \rho_w)g\nabla z\right] \quad (14)$$

# 3. Comsol validation using 1D Buckley-Leverett benchmark

Comsol Multiphysics (Comsol 2019) is a finite element-based software which allows to insert any kind of equation in the mathematics module whether it is algebraic, ordinary differential equations (ODE), partial differential equation (PDE) or a system or a mix of those.

Here Buckley-Leverett problem in 1D will be simulated using mathematics module to validate the usage of this procedure for our two-phase system in 3D simulation. Buckley-Leverett problem is a 1D problem that assumes water displacing oil as a piston-like process. Buckley-Leverett system can be derived from pressure and saturation equations (7) and (8):

$$\nabla \cdot u = q_o + q_w \quad (15)$$

$$\varphi \frac{\partial (S_w)}{\partial t} + u \frac{\partial f_w}{\partial x} = q_w \quad (16)$$

General form time-dependent PDE from mathematical module in Comsol is applied to pressure equation (15) with *p* as dependent variable. The components of the total velocity are the components of conservative flux vector. There is no source term, other than the injection and production wells. General form PDE in Comsol reads,

$$e_a \frac{\partial u}{\partial t} + d_a \frac{\partial u}{\partial t} + \nabla \cdot \boldsymbol{\Gamma} = f \quad (17)$$

Corresponds to the following coefficients,

$$\boldsymbol{\Gamma} = (u_x, u_y, u_z), \quad e_a = 0, \quad d_a = 0, \quad f = 0$$

With pressure as the dependent variable $\underline{\boldsymbol{u=p}}$ in equation (17).

The saturation equation (16) is applied in the coefficient form PDE which reads,

$$e_a \frac{\partial^2 u}{\partial t^2} + d_a \frac{\partial u}{\partial t} + \nabla \cdot (-c\nabla u - \alpha u + \boldsymbol{\gamma}) + \boldsymbol{\beta} \cdot \nabla u + au = f \qquad (18)$$

Corresponds to the following coefficients,

$$\boldsymbol{\gamma} = (u_{wx}, u_{wy}, u_{wz}), \quad e_a = 0, \quad d_a = \varphi, \quad \beta = 0, \alpha = 0, f = 0, a = 0, c = e$$

With water saturation as the dependent variable $\underline{\boldsymbol{u=S_w}}$ in equation (18).

The components of water, oil and total velocities are expressed as user-defined variables as listed in table 1.

Table 1. The components of velocity and saturation applied in Comsol as variables

| Variables | Description |
|---|---|
| $u_x = -k_x \lambda \, p_x$ | Total velocity, x component |
| $u_{wx} = f_w u_x$ | Water phase velocity, x component |
| $u_{ox} = f_o u_x$ | Oil phase velocity, x component |
| $S_o = 1 - S_w$ | Oil saturation |
| $f_o = \lambda_o / \lambda$ | Oil fraction flow |
| $f_w = \lambda_w / \lambda$ | Water fraction flow |
| $\lambda_o = k_{ro} / \mu_o$ | Oil phase mobility |
| $\lambda_w = k_{rw} / \mu_w$ | Water phase mobility |
| $\lambda = \lambda_w + \lambda_o$ | Total mobility |

The constant parameters of the model are defined in the parameters section and they are defined as in table 2. The numerical diffusivity $\underline{\boldsymbol{e}}$ is an additional isotropic term to stabilize the numerical solution. The permeabilities are defined as a function of the phase saturation,

$$k_{ro} = S_o^2$$

$$k_{rw} = S_w^2$$

Initial and boundary conditions for pressure and saturation equations are defined in table 3. For pressure equation, the Neumann boundary condition is specified as flux/source term at the inlet (injection well), constant pressure (Dirichlet) is fixed at the outlet (production well). For the water saturation equation, a fixed water saturation of 1 is defined at the inlet and a flux/source boundary condition in defined to represent outflow ($\nabla S_w = 0$).

Table 2. Constant parameters in the 1D model

| Parameter [unit] | Value | Description |
| --- | --- | --- |
| $k_x$ [mD] | 100 | Absolute permeability in x direction |
| $\varphi$ [-] | 0.3 | Porosity |
| $S_{wc} = S_{ini}$ [-] | 0.0 | Connate and initial water saturation |
| $S_{or}$ [-] | 0.0 | Residual oil saturation |
| $\rho_w$ [kg/m³] | 1000 | Density of water |
| $\rho_o$ [kg/m³] | 800 | Density of oil |
| $\mu_w$ [cP] | 1 | Viscosity of water |
| $\mu_o$ [cP] | 2 | Viscosity of oil |
| $u_{in}$ [m/s] | $1 \times 10^{-4}$ | Injection velocity |
| e [m²/s] | $5 \times 10^{-5}$ | Numerical Diffusivity |
| $p_{ini}$ [Pa] | 0 | Initial pressure in the 1D domain |
| $p_{out}$ [Pa] | 0 | Outlet pressure |

Table 3. Constant parameters in the 1D model

|  | Initial | Inlet | Outlet | Boundary |
|---|---|---|---|---|
| Pressure eq. | $p=p_{ini}$ | $u=u_{in}$ | $p=p_{out}$ | $-n.u=0$ |
| Saturation eq. | $S_w=S_{wc}$ | $S_w=1$ | $-n.(-c\nabla S_w + u_w)=\vec{u_w}$ | $-n.(-c\nabla S_w + u_w)=0$ |

The pressure and saturation are solved at each time step as dependent variables. The oil saturation is calculated from variable definition $S_o = 1-S_w$. The evolution of water saturation in 1m domain length at different time steps from 0 to 30 minutes is depicted in figure 1. It can be seen from this figure that the front represents a typical Buckley-Leverett displacement which justify using Comsol for immiscible two-phase flow.

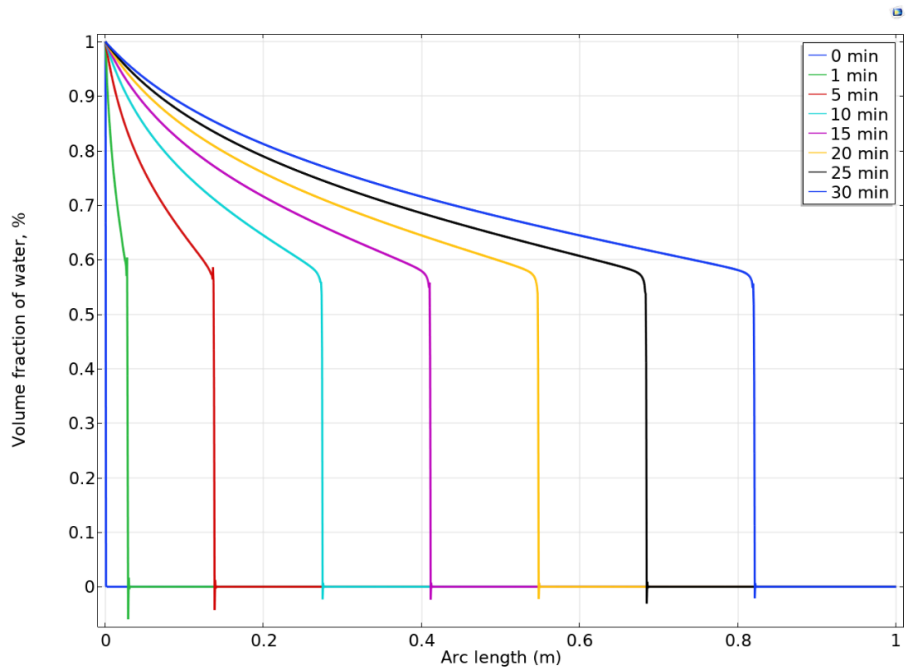

Figure 1. Displacement front at different time steps of Buckley-Leverett problem

## 4. Immiscible two-phase flow in 3D

Simulation steps using COMSOL Multiphysics® are the same as other commercial software but the sequence is different owing to the speciality in discretization procedure used in FEM, which is completely different from FDM-based commercial reservoir simulation packages. The flow of simulation in Comsol is in general as flows:

- Choose the domain dimension 0D, 1D, 2D or 3D
- Select the physics involved in the process which need to be simulated
- Choose steady state or transient simulation type
- Create the geometry
- Specify materials' properties
- Define initial and boundary conditions
- Create the mesh
- Run simulation
- Post-processing of the results

Mathematical module is used in this paper due to its merits to stabilize the solution. The general form and coefficient form PDE are used to solve pressure (p) and saturation (s) equations respectively. The numerical study is applied on the famous SPE case study quarter five-spot inverted model shown in figure 2 and meshed in figure 3. The chosen dimension is 100m*100m*6m with two wells located on the diagonal of the square-shaped domain.

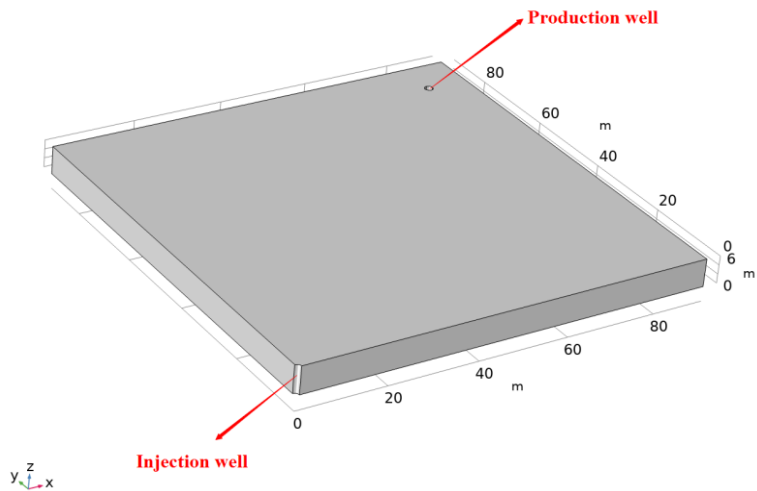

Figure 2. Quarter inverted five-spot model

All the parameters and variables in the problem are stated in tables 4 and 5 respectively. Relative permeability data are applied as analytic functions in the definitions node in Comsol and the values of permeability are tabulated in table 6.

The problem is simulated for 100 days and the CPU time is 7.46 minutes.

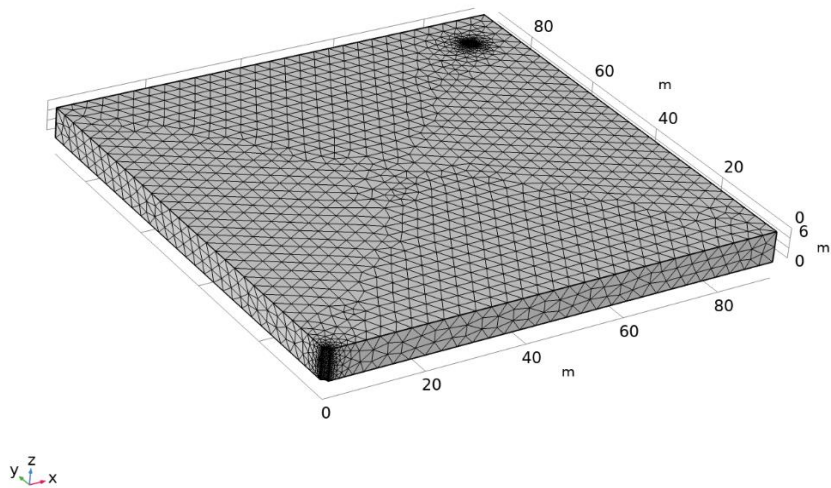

Figure 3. Triangular mesh of the domain

Table 4. The components of velocity and saturation applied in Comsol as variables

| Variables | Description |
|---|---|
| $u_x = -k_x \lambda\, p_x$ | Total velocity, x component |
| $u_y = -k_y\, p_y$ | Total velocity, y component |
| $u_z = -k_z(\lambda p_z - (\lambda_w \rho_w + \rho_o \lambda_o)g)$ | Total velocity, z component |
| $u_{wx} = f_w u_x$ | Water phase velocity, x component |
| $u_{wy} = f_w u_y$ | Water phase velocity, y component |
| $u_{wz} = f_w u_z + \lambda_o f_w k_z(\rho_w - \rho_o)g$ | Water phase velocity, z component |
| $u_{ox} = f_o u_x$ | Oil phase velocity, x component |
| $u_{oy} = f_o u_y$ | Oil phase velocity, y component |
| $u_{oz} = f_o u_z + \lambda_w f_o k_z(\rho_o - \rho_w)g$ | Oil phase velocity, z component |
| $S_o = 1 - S_w$ | Oil saturation |
| $f_o = \lambda_o/\lambda$ | Oil fraction flow |
| $f_w = \lambda_w/\lambda$ | Water fraction flow |
| $\lambda_o = k_{ro}/\mu_o$ | Oil phase mobility |
| $\lambda_w = k_{rw}/\mu_w$ | Water phase mobility |
| $\lambda = \lambda_w + \lambda_o$ | Total mobility |
| $p_{ini} = 1.0135*10^5 + \rho_w*g*(2453.64-z)$ | Initial pressure in the reservoir |
| $p_{out} = 1.0135*10^5 + (S_{wc}*\rho_w + (1-S_{wc})*\rho_o)*g*(2453.64-z)$ | Bottomhole pressure in production well |

Table 5. Constant parameters in the 3D model (Willhite 1986)

| Parameter [unit] | Value | Description |
|---|---|---|
| $k_x$ [m$^2$] | 250*10$^{-15}$ | Absolute permeability in x direction |
| $k_y$ [m$^2$] | 250*10$^{-15}$ | Absolute permeability in y direction |
| $k_z$ [m$^2$] | 25*10$^{-15}$ | Absolute permeability in z direction |
| $\varphi$ [-] | 0.15 | Porosity |
| $S_{wc} = S_{ini}$ [-] | 0.363 | Connate and initial water saturation |
| $S_{or}$ [-] | 0.205 | Residual oil saturation |
| $\rho_w$ [kg/m$^3$] | 1000 | Density of water |
| $\rho_o$ [kg/m$^3$] | 850 | Density of oil |
| $\mu_w$ [Pa.s] | 0.001 | Viscosity of water |
| $\mu_o$ [Pa.s] | 0.002 | Viscosity of oil |
| $u_{in}$ [m/s] | 5*10$^{-5}$ | Injection velocity |
| e [m$^2$/s] | 5*10$^{-5}$ | Numerical Diffusivity |

Table 6. Relative permeability data (Willhite 2018)

| $S_w$ | $k_{rw}$ | $k_{ro}$ | $S_w$ | $k_{rw}$ | $k_{ro}$ |
|---|---|---|---|---|---|
| 0.363 | 0 | 1 | 0.58 | 0.060 | 0.168 |
| 0.380 | 0 | 0.902 | 0.6 | 0.084 | 0.131 |
| 0.400 | 0 | 0.795 | 0.62 | 0.113 | 0.099 |
| 0.420 | 0 | 0.696 | 0.64 | 0.149 | 0.073 |
| 0.440 | 0.001 | 0.605 | 0.66 | 0.194 | 0.051 |
| 0.460 | 0.003 | 0.522 | 0.68 | 0.247 | 0.034 |
| 0.480 | 0.006 | 0.445 | 0.7 | 0.310 | 0.021 |

| 0.5   | 0.011 | 0.377 | 0.72  | 0.384 | 0.011 |
| 0.520 | 0.018 | 0.315 | 0.74  | 0.470 | 0.005 |
| 0.54  | 0.028 | 0.260 | 0.76  | 0.57  | 0.002 |
| 0.56  | 0.042 | 0.210 | 0.795 | 0.78  | 0     |

## 5. Discussion

The simulation results are illustrated in figures 4 through 10. The oil saturation is calculated from variable definition $S_o = 1 - S_w$. The evolution of phase saturation at 20 and 60 days of the water injection are depicted in figure 4. After 40 days of injection, water has reached the production well. The gradual colour change in the saturation figures indicates stable solution (there are no anomalies in colours in the isosurfaces of saturation).

Figures 5 and 6 show the phase saturation values in the diagonal plane of the reservoir from the injection well to the production well. The results are consistent with the relative permeability curves and its end-points. The steep saturation curves show that the water flooding is convection dominant. The average saturation in the reservoir is shown in figure 7. The average saturation of the oil is decreasing linearly until the water breakthrough. The fractional flow curves for water and oil are presented figure 8. It can be seen from this figure that as oil fraction decreases, water fraction increases. As can be seen, the breakthrough time occurs around $40^{th}$ day of injection. At the end of injection period, the well is producing only water, where $f_w$ has reached 1.

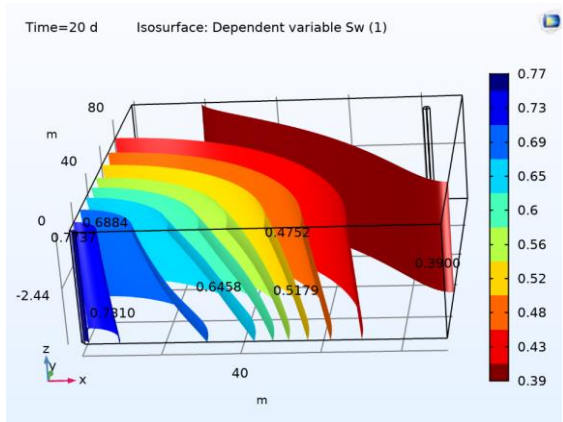 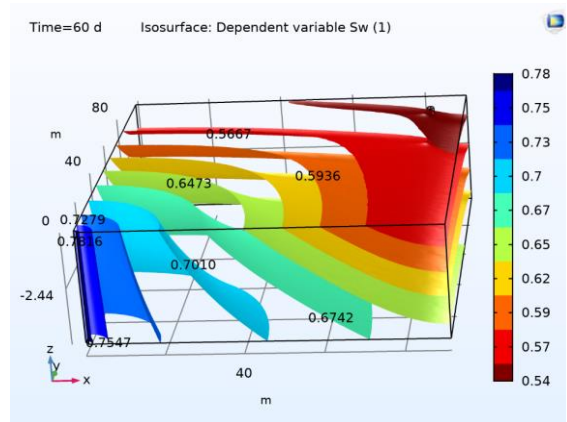

(a) Water saturation, $S_w$ after 20 days  (b) Water saturation, $S_w$ after 60 days

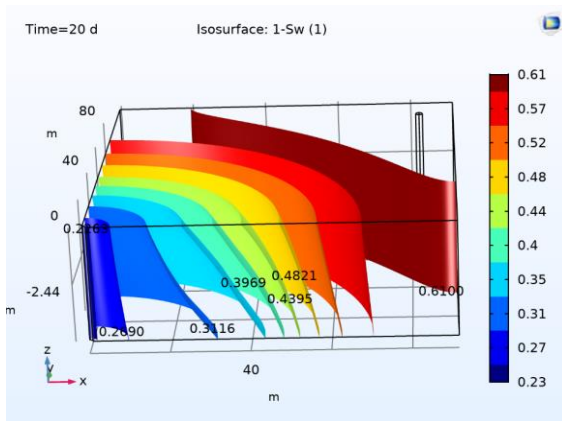 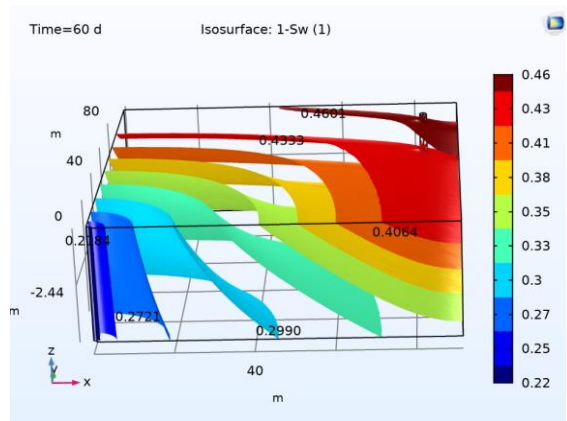

(c) Oil saturation, $S_o$ after 20 days  (d) Oil saturation, $S_o$ after 60 days

Figure 4. Water and oil saturation isosurfaces at 20 and 60 days (vertical scale of the domain is multiplied by 5 for better view only)

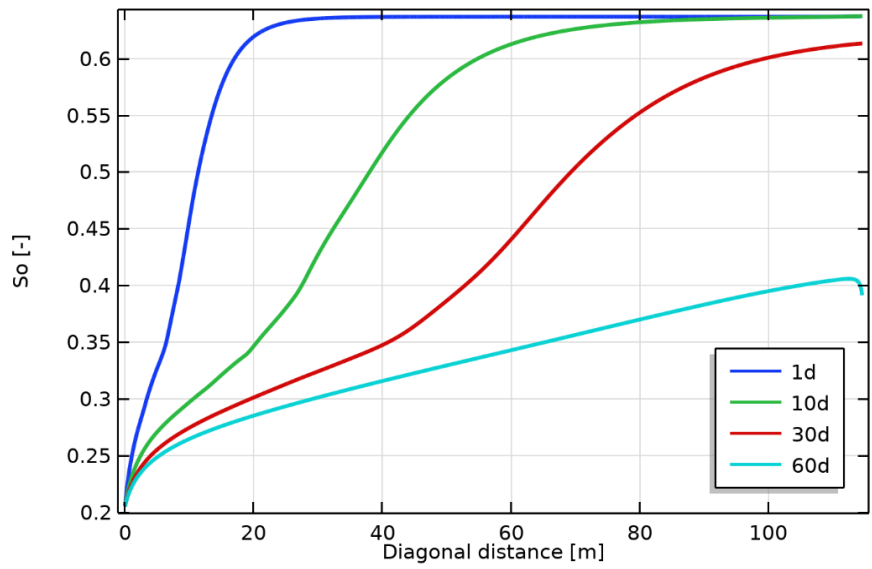

Figure 5. Oil saturation in the diagonal plane from the injection to the production well

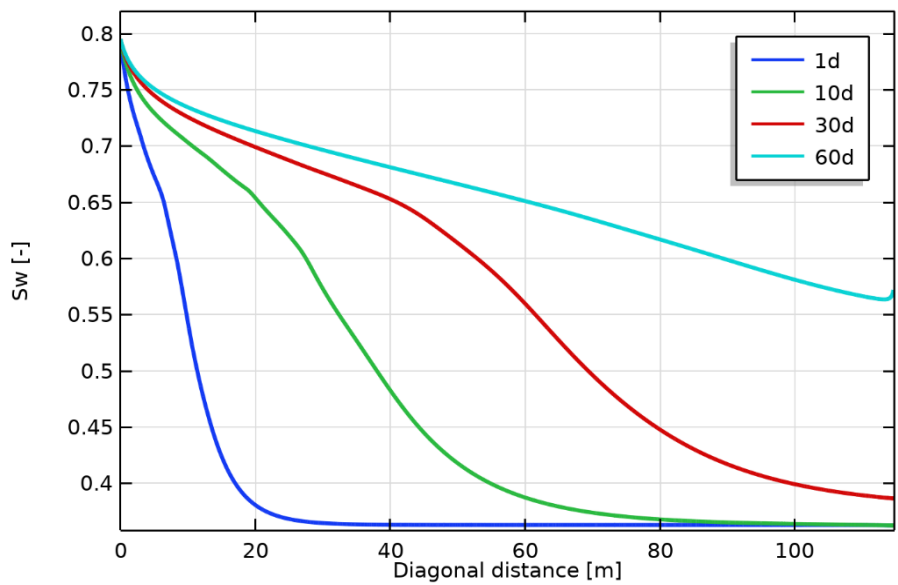

Figure 6. Water saturation in the diagonal plane from the injection to the production well

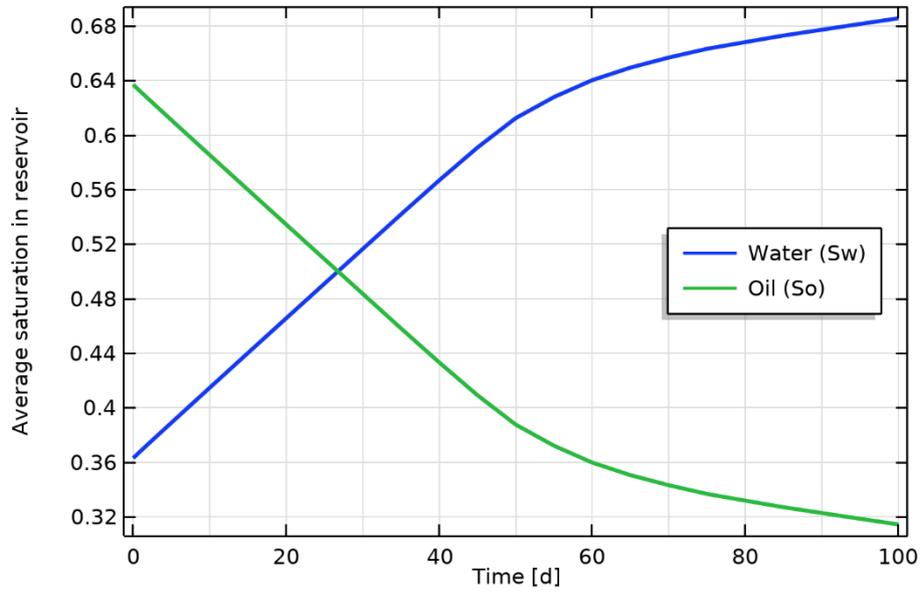

Figure 7. Average saturation in the reservoir

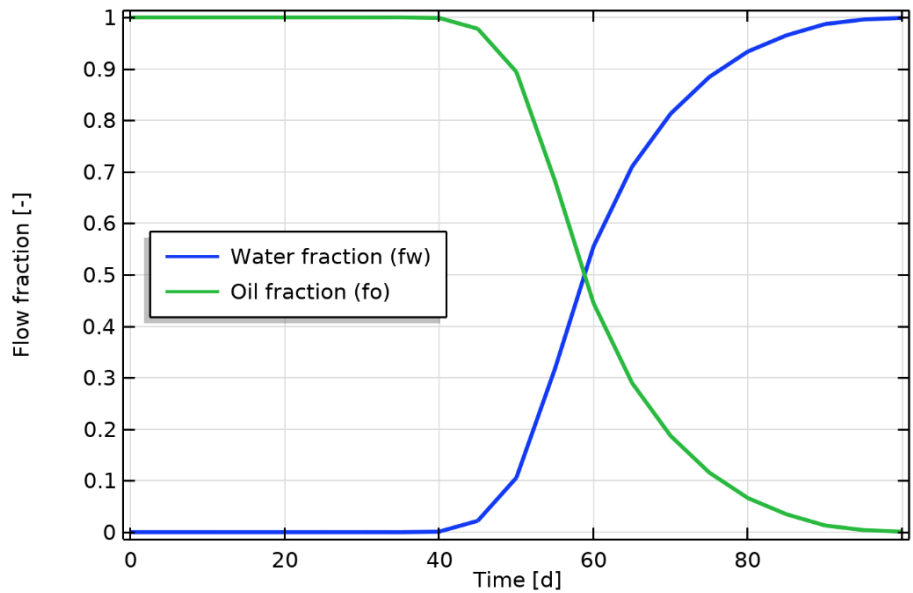

Figure 8. Fractional flow curves at the production well

Figure 9 depicts the change of bottomhole pressures at the injection and production well with time. The initial pressure in the reservoir is 241 bar. The injection pressure increases to 245 bar at first day and stays constant 20 days. The gradual injection pressure increases from 40-50 days and the decrease afterwards indicates the water breakthrough at the production well.

As can be seen from figure 10, the daily oil production of 238 bbl/d decreases exponentially after breakthrough to 0 just few days before the 100$^{th}$ day due to the water breakthrough.

The qualitative assessment of the simulation results show that the two-phase flow equations were implemented correctly in the water flooding model.

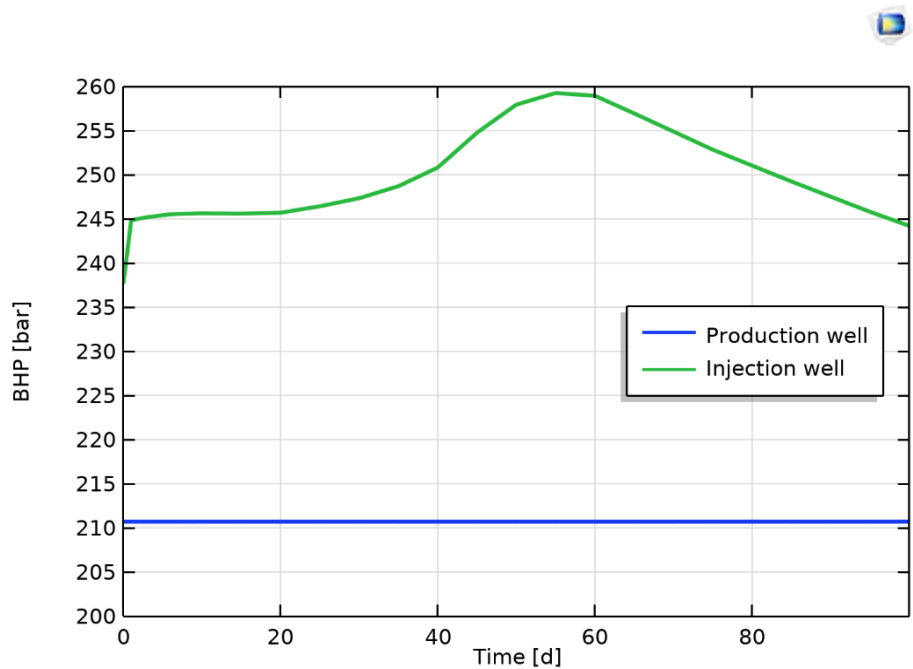

Figure 9. Bottomhole pressure at injection and production wells

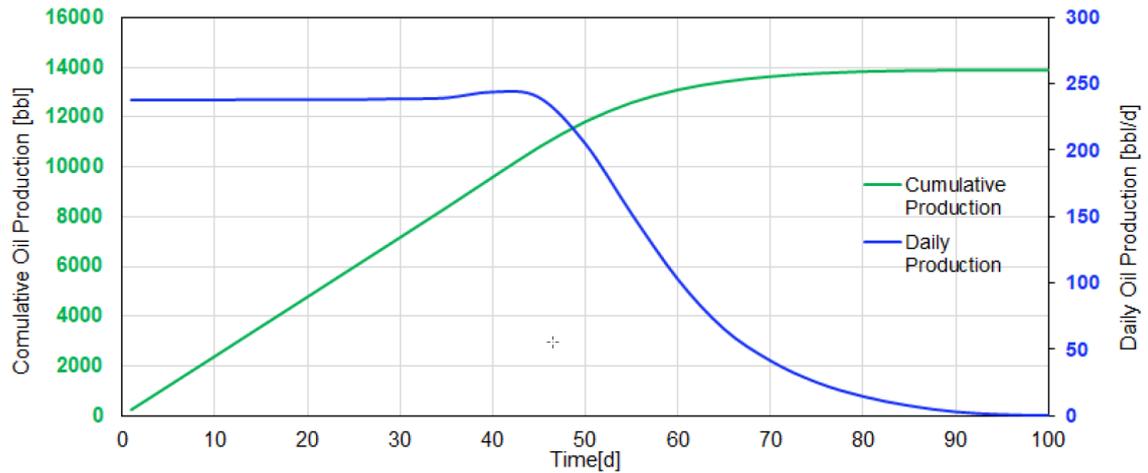

Figure 10. Oil production vs. Time

## 6. Conclusion

In this paper, finite-element method (FEM) is introduced as a numerical tool to simulate an example of immiscible two-phase flow in 3D and prior to that, Buckley-Leverett problem is used to validate using the FEM- based software introduced in this paper. To achieve the proposed targets, the mathematical model of the problem is reviewed and formulated in phase form to be applied easily in the mathematical module of Comsol Multiphysics. The results show that the FEM can be used efficiently to simulate the two-phase flow in oil reservoir with relatively an acceptable CPU time.

## Acknowledgments

The first author would like to acknowledge German Academic Exchange Service (DAAD) and Leadership for Syria program (LfS) for sponsoring his PhD research scholarship at Freiberg University of Technology.

**Nomenclature**

g    gravity acceleration, m/s$^2$

k    absolute permeability, m$^2$

$k_{row}$    two-phase relative permeability to oil in oil–water system

$k_{rwo}$  two-phase relative permeability to water in oil–water system

p     pressure, Pa

$p_c$   capillary pressure, Pa

$S_w$   water saturation

$S_o$   oil saturation

$S_{ro}$   residual oil saturation

$S_{wi}$  connate water saturation

t    time, s

u    dependent variable in equations(17) and (18)

$u_\alpha$   phase velocity, m/s

o     oil

w    water

x, y, z    cartesian coordinate axes

**Greek letters**

$\Gamma$  flux component

$\nabla$  del operator

$\alpha$  fluid phase notation

$\lambda$  mobility, m.s/kg

$\mu$  viscosity, Pa.s

$\rho$  density, kg/m³

$\varphi$  porosity, %

$\partial$  partial derivative notation